\documentclass[aps,pra,reprint,superscriptaddress,twocolumn]{revtex4-1}

\usepackage{amsmath}
\usepackage{amsfonts}
\usepackage{braket}
\usepackage{graphicx}
\usepackage{dsfont}
\usepackage{float}

\usepackage{xcolor}
\usepackage{soul}
\usepackage[normalem]{ulem}

\usepackage{natbib}
\usepackage{hyperref}
\begin{document}

\title{Measuring average of non-Hermitian operator with weak value in a Mach-Zehnder interferometer}

\author{Gaurav Nirala}
\altaffiliation{Homer L. Dodge Department of Physics and Astronomy,
The University of Oklahoma, Norman, OK, 73019, USA}
\affiliation{Light and Matter Physics, Raman Research Institute, Bengaluru-560080, India}

\author{Surya Narayan Sahoo}
\affiliation{Light and Matter Physics, Raman Research Institute, Bengaluru-560080, India}

\author{Arun K Pati}
\affiliation{Quantum Information and Computation Group, Harish-Chandra Research Institute, HBNI, Allahabad 211019, India}

\author{Urbasi Sinha}
\email[]{usinha@rri.res.in}
\affiliation{Light and Matter Physics, Raman Research Institute, Bengaluru-560080, India}

\begin{abstract}
Quantum theory allows direct measurement of the average of a non-Hermitian
operator using the weak value of the positive semidefinite
part of the non-Hermitian operator. Here, we experimentally demonstrate  the
measurement of weak value and average of non-Hermitian operators by a novel interferometric technique. Our scheme is unique as we can directly obtain the weak value from the interference visibility and the phase shift in a Mach Zehnder interferometer without using any weak measurement or post selection. Both the experiments discussed here were performed with laser sources, but the results would be the same with average statistics of single photon experiments. Thus, the present experiment opens up the novel possibility of measuring weak value and the average value of non-Hermitian operator without weak interaction and post-selection, which can have several technological applications.
\end{abstract}

\maketitle

\section{INTRODUCTION}

The representation of observables in quantum mechanics has been postulated to be restricted to Hermitian operators \cite{Griffiths}. However, it has been shown that the expectation value of a non-Hermitian operator can be inferred by measuring the weak value of the Hermitian operator into which the non-Hermitian operator can be polar decomposed \cite{nHP}. Weak Measurements and weak values have not only found technological applications in ultra sensitive measurements \cite{Dixon} but also in exploring foundational issues in quantum mechanics \cite{UncertaintyWeak,Lundeen}. In this manuscript, the novel idea regarding application of weak value to obtain expectation value of non-Hermitian operator given in Ref. \cite{nHP} is furthered and experimentally implemented.

Experimentally realizable outcomes are described by real numbers  \cite{Dirac} and in quantum mechanics, we demand that all observables be represented by Hermitian operators since their eigenvalues are real. However, it has been argued that, demanding a) eigenvalues to be real and  b) validity of Spectral theorem (existence of complete orthonormal eigenbasis), the general class of operators that may be used to describe observables are normal operators \cite{Zhang}. Thus demanding Hermiticity is a sufficient condition for eigenvalues to be real but not necessary. For a generic non-Hermitian operator, the eigenstates need not be orthogonal and hence experimentally may not be distinguishable. Therefore, the average value of a non-Hermitian operator cannot be obtained from statistics of outcomes. This is another fundamental problem in measuring the expectation value of a non-Hermitian operator apart from the fact that it is usually complex. From an experimental perspective, a complex number can be said to have been measured if, with the same experimental setup, we can individually measure its components either decomposed into real and imaginary parts or as magnitude and phase. Since weak values are in general complex and weak measurements do not require distinguishing between the eigenstates of the operator, expectation value of the non-Hermitian operator can be inferred by expressing it in terms of weak value of the positive semi-definite part of the non-Hermitian operator. 


Here, we experimentally demonstrate a novel interferometric scheme whereby we can infer weak values without performing any weak measurements. Thus, not only do we circumvent dealing with the weakness criteria but also avoid the need to have post-selection performed.

This article is structured as follows. We begin with a description of weak value as a result of weak measurement and then explain how using a Mach-Zehnder interferometer we can obtain not only the weak value but also infer the expectation value of non-Hermitian operators. Finally, we compare the results and inferences of weak value from experiments performed using a Mach-Zehnder interferometer with  conventional weak measurement experiments. The former involves obtaining the general complex expectation value of a non-Hermitian operator in terms of the complex magnitude and phase in the Argand plane. Other implementations may involve separately obtaining the real and imaginary part or reconstructing the expectation value from the Hermitian traceless unitary basis (for e.g. the Pauli basis for 2 dimensional systems) of the operator space. Finally, we conclude the paper with discussions and summary.

\subsection{Weak Values and Weak Measurements}
In quantum mechanics, the result of measurement performed on any spin component of a particle cannot be predicted with certainty when the ensemble is prepared in a state which is not the eigenstate of the operator corresponding to the observable of the spin component. Aharonov, Albert and Vaidman(AAV) envisioned that to assign a unique value to any spin component at time $t$, the knowledge of the initially prepared state alone is not sufficient. They argued that we would need the result of two measurements, one performed before $t$ and one performed after $t$ \cite{AAV1987}. The state prepared before time $t$ which, in general, is the result of the evolution of the selected state out of many outcomes of a measurement performed before $t$ is called the pre-selected state $\ket{\psi}$. The second measurement performed after $t$ can have multiple outcomes and any subset of the outcomes can be selected. This procedure is called post-selection. The state just after time $t$ whose evolution guarantees a successful post-selection is called the post-selected state $\ket{\phi}$. This unique value, however, cannot be the result of a conventional strong measurement where the state reduces to one of the eigenstates. Later, a method to realize this unique value was described by AAV which is now known as weak-measurement \citep{AAV1988}. This unique value obtained as a result of weak measurement is known as the weak value.

The effective interaction Hamiltonian in a von-Neumann measurement process of an observable described by $S$ can be written as
\begin{align}
H = g(t) S \otimes P_{x} ,
\end{align}
where $g(t)$ is a compact function of time which is non-zero during the interaction. Here, $P_x$ is the conjugate momentum corresponding to a pointer variable $x$. After the interaction is over, any initial distribution of pointer states gets displaced by $a$. If the displacement is much much larger than the initial uncertainty $\sigma$ of the distribution of $x$, then the measurement outcomes can be distinguished and the interaction is said to be strong. However, if the displacement caused is smaller than the uncertainty of pointer state distribution, the system state is said to have not reduced.  Further, post-selection allows interference which may result in a large shift of the mean of the pointer. This shift is proportional to the weak value given by
\begin{align}
{}_{\phi}\braket{S}_{\psi}^{(w)} = \frac{\bra{\phi}S\ket{\psi}}{\braket{\phi|\psi}} .
\label{Eqn_weakValue}
\end{align}

The weak value can be complex and can lie outside the eigenspectrum. This was experimentally realized soon after the prediction \cite{Ritchie}. Since then weak measurements have been used in foundational studies of quantum mechanics like Hardy's paradox \cite{Hardy} and average Bohmian \cite{Bohmian} trajectories as well as expanded the field of mathematical properties of super-oscillations \cite{Superoscillation}.

\subsection{Expectation value of non-Hermitian Operators with Weak Value }

Consider the operator $A$, which in general can be non-Hermitian. The expectation value of $A$ in state $\ket{\psi}$ could in general be a complex number $z=\braket{\psi|A|\psi}$. In this section, we shall discuss how polar decomposition can be used to recast the complex expectation value of any non-Hermitian operator in terms of a complex weak value of a Hermitian operator. The complex weak value of a Hermitian operator can be experimentally obtained by measuring its real part from the shift of the pointer variable and its imaginary part inferred from the shift of the momentum conjugate to the pointer variable.

Given any operator $A$, we can always polar decompose \cite{Hall} as  $A = U R$, where $U$ is unitary operator and $R$ is the Hermitian semi-definite operator obtained as $R = \sqrt{A^\dagger A}$. 

Following the idea presented in Ref. \cite{nHP}, the expectation value of A can now be expressed in terms of the weak value of $R$ as follows
\begin{align}
z&=\braket{\psi|A|\psi}  \label{z def}\\ 
&=\braket{\psi|U R|\psi}  \label{z expression}\\
&=\frac{\braket{\psi|U R|\psi}}{\braket{\psi|U|\psi}} \braket{\psi|U|\psi}\\ 
&=\frac{\braket{\phi| R|\psi}}{\braket{\phi|\psi}} \braket{\phi|\psi}\\  
&= {}_\phi \braket{R}_\psi^{(w)} \braket{\phi|\psi} .
\end{align}
Here, $\ket{\phi} = U^\dagger\ket{\psi}$. Given the polar decomposition, we can experimentally measure the weak value of $R$ in the pre-selected state $\ket{\psi}$ and the post-selected state $\ket{\phi}$. The post-selected state is uniquely determined by the unitary $U$ which in turn is unique for a given $R$. Thus, the expectation value of a non-Hermitian operator can be restated as the transition amplitude mediated by the Hermitian polar component $R$ from the state $\ket{\psi}$ in which the expectation value is to be measured to a state $\ket{\phi}$ into which $\ket{\psi}$ can evolve into with the unitary polar component $U$. Note that $\braket{\phi|\psi}$ in general can be complex but since $\ket{\psi}$ and $\ket{\phi}$ will be the pre- and post-selected state in the weak measurement of $R$ and hence can be always computed. Knowing the weak value of $R$ and the expectation value of $U$ in $\ket{\psi}$, i.e., $\braket{\psi|U|\psi}$, we can infer the complex expectation value $z=\braket{A}$.

Conversely, if by some other method, we obtain the expectation value of $A$  and also the expectation value of $U$, we can infer the weak value of $R$. Then, we can experimentally obtain weak value without performing weak measurement. Such a method which involves Mach-Zehnder interferometry is described in the next section. Although the method uses optics as an example, but is applicable, in general, to all other quantum systems.

\section{Average value of non-Hermitian Operator using MACH-ZEHNDER INTERFEROMETER}
\begin{figure}[H]
\includegraphics[width=\linewidth]{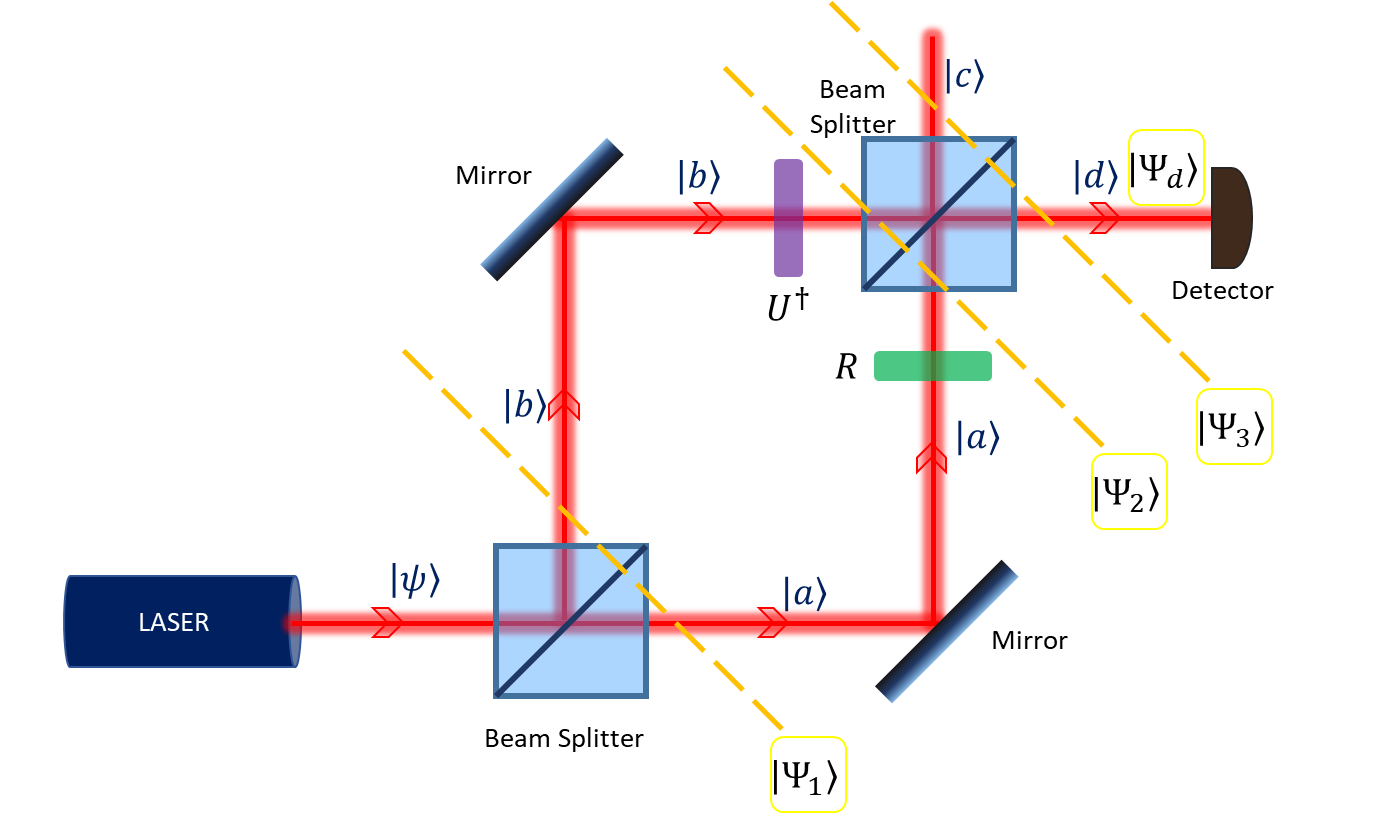}
\caption{Schematic for Mach-Zehnder Interferomer. Optical components $R$ is placed in one arm and $U^\dagger$ in the other.}
\label{fig1_MZ2D}
\end{figure}

The interferometric scheme conceptualized by Zehnder \cite{Zehnder} and later enhanced by Mach \cite{Mach} has not only found applications in engineering optical devices \cite{Ohmae:11}  but also is used in experiments concerning quantum foundations \cite{Vincent,Bomb}. In this section, we describe a new scheme by which we can infer the expectation value of a non-Hermitian operator $A$ using the Mach-Zehnder interferometer (see Figure \ref{fig1_MZ2D}).

In experiments, since we directly measure real quantities only, the complex expectation value can be inferred from two real quantities obtained from the experiment. One way to infer complex $z= |z| \exp(i \varphi)$ is to measure the magnitude $|z|$ and phase $\varphi$ independently. This can be achieved by measuring the visibility and the phase shift in a Mach-Zehnder interferometer, which can have optical elements corresponding to operators into which $A$ can be polar decomposed.

We shall consider $A$ to be any non-Hermitian operator affecting the polarization degree of freedom. The only requirement is that the operators $U$ and $R$ when represented in Jones matrix formalism must describe optical components which can be realized in a laboratory.

Consider the state $\ket{\psi}$ to be the polarization state of the input beam to the Mach Zehnder interferometer. The state after the first 50:50 Hadamard type beam splitter \cite{Hadamard} would be given by 
\begin{align}
\ket{\Psi_1} = \frac{1}{\sqrt{2}}(\ket{a}+\ket{b}) \ket{\psi},
\end{align}
where $\ket{a}$ and $\ket{b}$ are spatial modes corresponding to arm $a$ and arm $b$ respectively as described in Figure  \ref{fig1_MZ2D}.  Now, we place the optical components corresponding to the operators $R$ in arm $a$ and $U^\dagger$ in arm $b$ which act on polarization degree of freedom. Note that if $U$ is experimentally realizable, so is $U^\dagger$. The evolved state $\ket{\Psi_2}$, just before the second beam splitter, is given by
\begin{align}
\ket{\Psi_2}=\frac{1}{\sqrt{2}}(R \ket{\psi} \ket{a}+ U^\dagger \ket{\psi}\ket{b}) .
\end{align}
When the two beams finally recombine at the second beam splitter, there would be a relative phase difference between the two arms owing to phase changes due to propagation and reflections, which can be denoted as $\epsilon$. Note that $\epsilon$ is not due to the operators $R$ and $U^\dagger$, but includes the phase difference caused due to path difference between the two arms which includes differences in material refractive index and thickness of optical components.
After the second beam splitter, the state described in terms of ports $\ket{c}$ and $\ket{d}$, is given by
\begin{multline}
\ket{\Psi_3} = 
\left(\frac{1}{2}(R \ket{\psi}- U^\dagger \exp(i \epsilon)\ket{\psi})\right) \ket{c}\\
+\left(\frac{1}{2}(R \ket{\psi}+ U^\dagger \exp(i \epsilon)\ket{\psi})\right) \ket{d} .
\end{multline}
The detector placed in port $\ket{d}$ only detects the
component of the total state in the detector arm of the final beam splitter. This can be obtained by acting the projector $\Pi_d = \ket{d}\bra{d}$ on the entire state. The component of the state in the detector arm then becomes
\begin{align}
\ket{\Psi_d}= \Pi_d \ket{\Psi_3} = \frac{1}{2}(R \ket{\psi}+ U^\dagger \exp(i \epsilon)\ket{\psi}) \ket{d} .
\end{align}
Note that the above component depends of the phase shift $\epsilon$ between the two arms $\ket{a}$ and $\ket{b}$ along with the operators $R$ and $U$.

The intensity at the detector port of the final beam splitter is given by
\begin{align}
I_d(\epsilon) &= |\braket{d|\Psi_3}|^2 = ||\Psi_d||^2 \\
& = \frac{1}{4} (\braket{\psi|R^\dagger R|\psi}+ \braket{\psi|U U^\dagger |\psi} + \\
& \exp(i \epsilon)\braket{\psi|R^\dagger U^\dagger|\psi} +\exp(-i \epsilon)\braket{\psi|U R|\psi} ) .
\end{align}

Since $R$ is Hermitian and $U$ is unitary, the first term is the expectation value of $R^2$, and hence is real, while the second term is 1 as $U U^\dagger  = \mathds{1}$. The last two terms are in general complex and are the conjugate of each other. 


We finally have the intensity at the detector expressed in terms of $z$ as
\begin{align}
I_d = \frac{1}{4}\left(1+\braket{R^2}+2 |z| \cos(\varphi-\epsilon) \right) ,
\label{vis_expression}
\end{align}
where $\phi = \arg(z)$.

We have visibility as given by
\begin{align}
V &= 
\frac{\max{(I_d)} - \min{(I_d)}}{\max{(I_d)}+\min{(I_d)}} \\
& = \frac{2 |z|}{1+\braket{R^2}} .
\label{Vis_Avg}
\end{align}

Experimentally, if we obtain intensity $I_d$ vs $\epsilon$, say by varying optical path difference between the two arms of the Mach Zehnder, we can measure $V$ as the visibility and $\varphi$ as the phase shift caused due to the action of $R$ and $U$. Then, we need to determine $\braket{R^2}$ to determine $|z|$ from the knowledge of $V$. Since $R$ is Hermitian, we have $\braket{R^2} = ||R\psi||^2$. Thus the expectation value of $R^2$, can be experimentally obtained by measuring the power throughput after passing a beam with the state $\ket{\psi}$ through the operator $R$.

As an example, we consider the the spin lowering ladder operator defined for arbitrary spin systems as 
$
\sigma^+_- = \frac{1}{2}(\sigma_x \mp i \sigma_y)
$
.
In the Jones matrix representation, we have
\begin{align}
A =
\begin{pmatrix}
0 & 0 \\
1 & 0
\end{pmatrix} 
=
 \begin{pmatrix}
0 & 1 \\
1 & 0
\end{pmatrix} 
\begin{pmatrix}
1 & 0 \\
0 & 0
\end{pmatrix}
\label{lowering operator} 
\end{align}
The unitary polar component $U = \sigma_x$ 
, can be realized as a half wave plate with fast axis rotated $45 ^\circ$ from horizontal. The Hermitian operator $ R = \ket{H} \bra{H}$, can be realized as a Polariser with transmission axis set to Horizontal (or by considering the transmitted arm of the Polarizing beam splitter (PBS), where we neglect the reflected beam).

\section{Experimentally obtaining Visibility}

\begin{figure}[H]
\includegraphics[width=\linewidth]{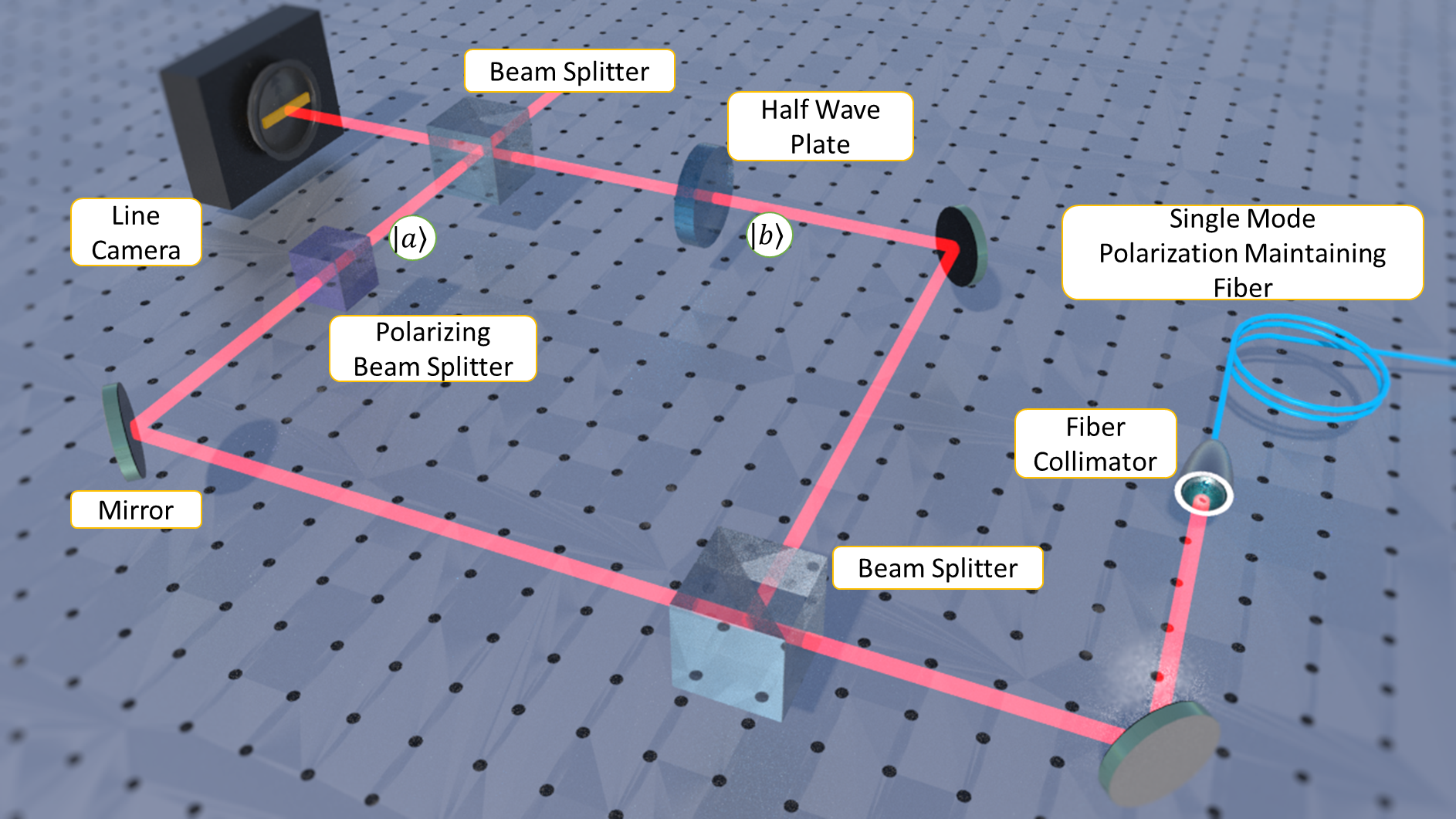}
\caption{Experimental Mach Zehnder Setup: We have Polarizing beam splitter(PBS)/polarizer oriented along Horizontal in one arm. To compensate for elliptic polarization components after reflection from first beam splitter, we place a Quarter wave-plate ((QWP) before the Half wave-plate (HWP).}
\label{fig2_MZ3D}
\end{figure}

We use Ti-sapphire laser continuous wave single mode fibre output at $810 (\pm 2)$ nm and pass it through a polarizing beam splitter (PBS)  to make the polarization linear. Then we use a half wave plate(HWP) to make the polarization state $\ket{\psi}$ input to the Mach Zehnder as diagonally polarized state $\ket{+} = \frac{1}{\sqrt{2}}(\ket{H}+\ket{V})$. The HWP was rotated to make the transmission and reflected power of the PBS placed in one arm of the interferometer equal. Thus, by virtue of alignment procedure, we verify that

\begin{align}
\ket{\psi} = \ket{+} \Rightarrow \braket{\psi|R^2|\psi} = \frac{1}{2}
\end{align}

One can controllably vary path difference to obtain $I_d$ as a function of $\epsilon$ so that $|z|$ and phase shift $\varphi$ can be obtained. But, for our choice of the non-Hermitian operator and the $\ket{\psi}$, the expectation value turns out to be real i.e. ,

\begin{align}
\frac{1}{\sqrt{2}}
\begin{pmatrix}
1 & 1
\end{pmatrix}
\begin{pmatrix}
1 & 0 \\
0 & 0
\end{pmatrix}
\frac{1}{\sqrt{2}}
\begin{pmatrix}
1 \\
1
\end{pmatrix}
= \frac{1}{2}
\end{align}

At this point, it is worth mentioning that indeed our method is
applicable in general to obtain a complex weak value but our aim here is to prove the efficacy of this novel method so we did not do the otherwise necessary phase stabilization.

\begin{figure}[H]
\vspace{30pt}
\includegraphics[width=\linewidth]{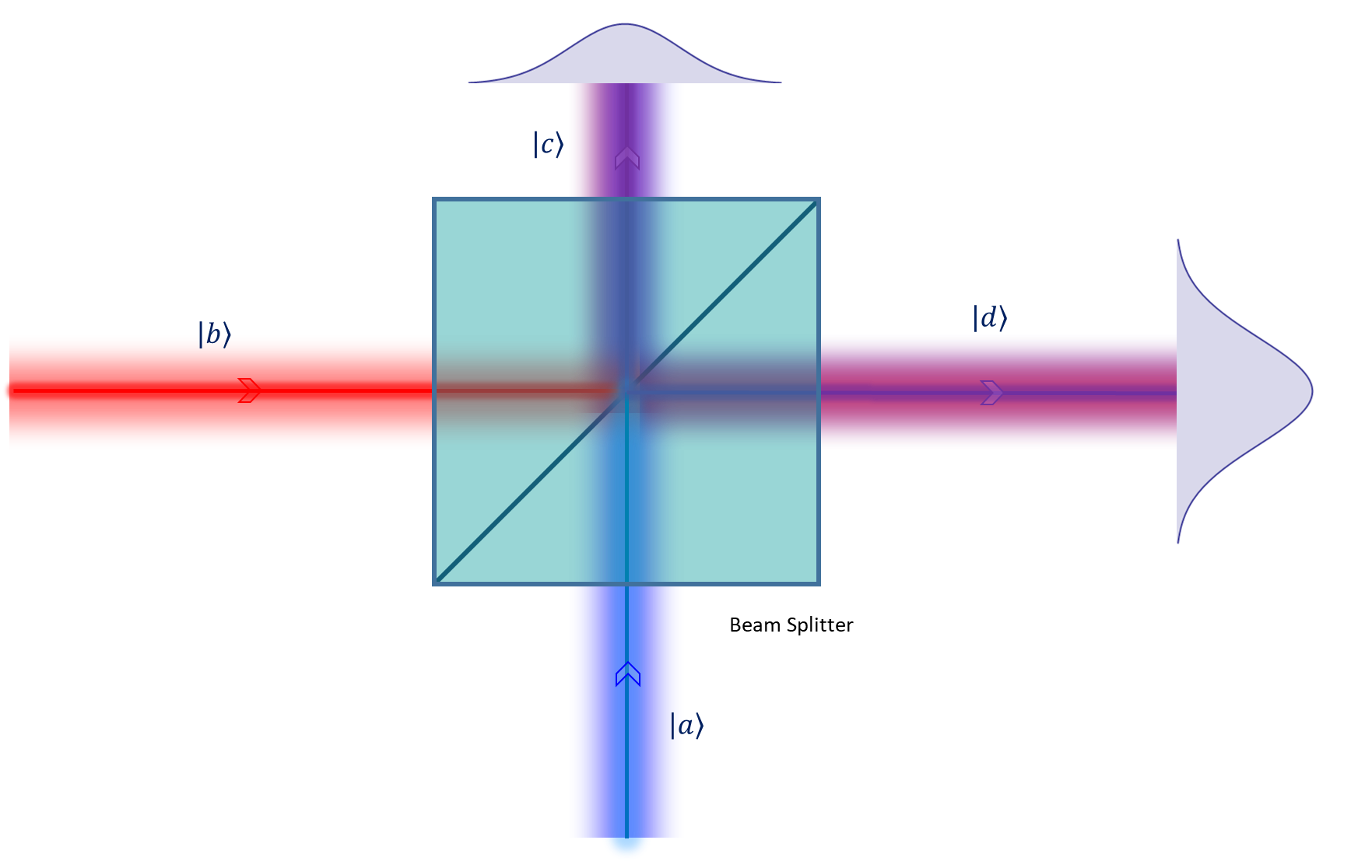}
\includegraphics[width=\linewidth]{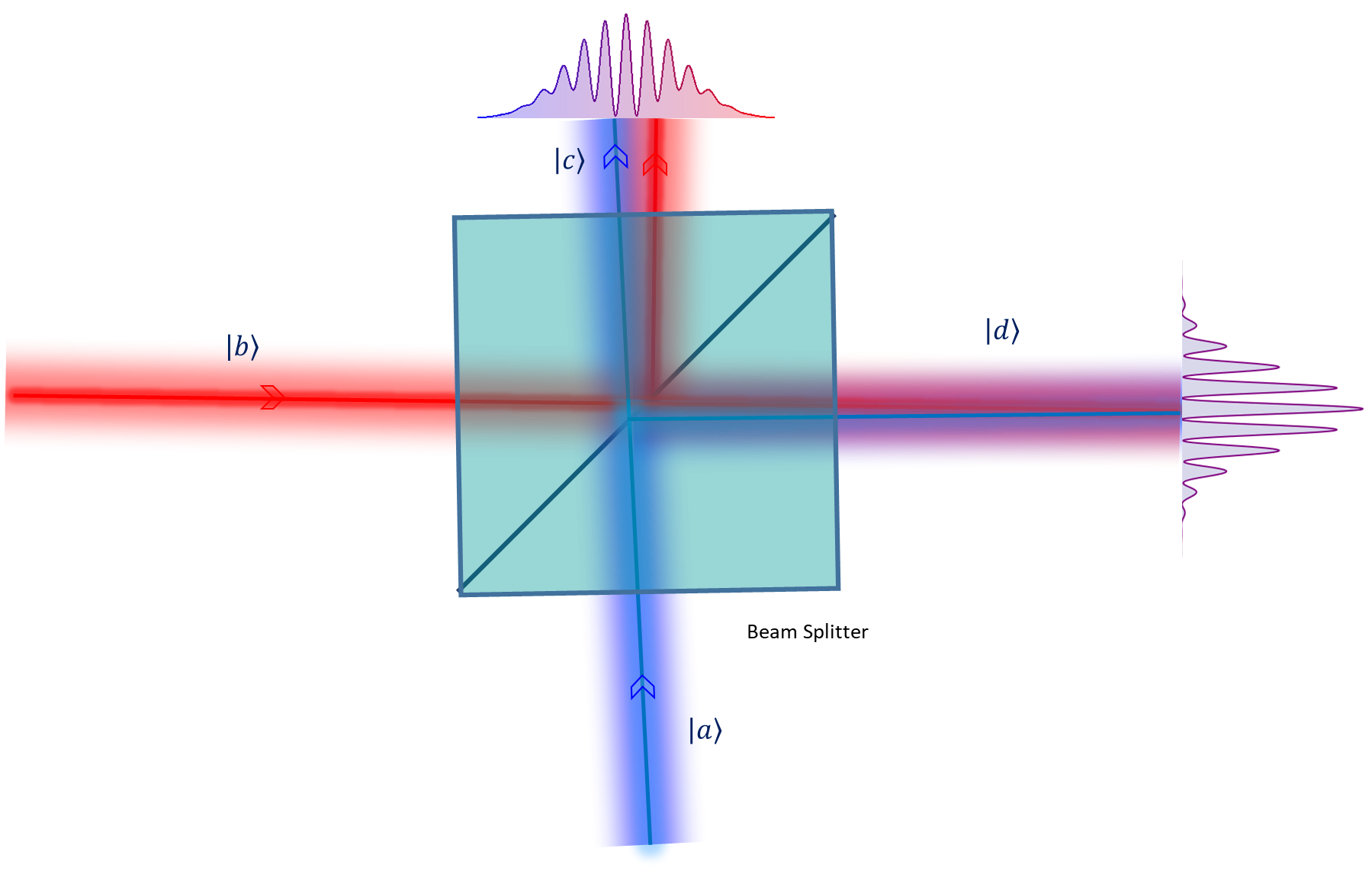}
\caption{Collinear (top) vs Non-Collinear configuration (bottom) of the Mach Zehnder:Consider the above beam splitter to be the second beam splitter of the Mach Zehnder Interferometer shown in experimental schematic Figure \ref{fig2_MZ3D}.}
\label{fig_Coliinearity}
\end{figure}

Since, we already know $\braket{R^2}$, if we can directly know visibility $V$, we can obtain the expectation value of the non-Hermitian operator $|z|$ using Eq. (\ref{Vis_Avg}).
To experimentally obtain the visibility, we use an easier technique, i.e., to align the Mach Zehnder interferometer in a non-collinear configuration, with a slight angle between the two beams incident at the final beam splitter, so that we see spatial fringes \cite{NonCollinear}. 
This is attained by first aligning the interferometer in an almost collinear configuration and then iteratively tweaking the mirrors to create a small displacement between the two beams incident at the final beam splitter. 
Then the final beam splitter is tilted to make the two beams overlap at the detector so that we have spatial fringes similar to a double slit interference pattern. Along the transverse axis of the fringes, we obtain $I_d$ vs. $\epsilon$ as the path taken by two beams coming from two arms to reach the same detector point varies within the Gaussian envelope.

In the collinear configuration (Figure \ref{fig_Coliinearity}  top) the beams incident from port $\ket{a}$ and $\ket{b}$ emerge along the same direction along port $\ket{c}$ and $\ket{d}$. The beam shape in the exit port remains Gaussian if the incident beams are Gaussian and the point of incidence on the beam splitter for beams coming from both the input ports is same. Only the intensity in the output ports vary depending on the phase difference $\epsilon$. In non-collinear configuration (Figure \ref{fig_Coliinearity} bottom), the two incident beam have a slight angle between them. If the point of incidence for beams coming from port $\ket{a}$ and port $\ket{b}$ are slightly separated, then the beam splitter can be tilted slightly so that on one exit port (say $\ket{c}$) the beams diverge and on the other exit port (say $\ket{d}$) the beams converge.

 Since the interferometer was not phase stabilized, $\epsilon$ may have random zero errors. Given an interference profile, we can obtain the visibility by knowing the individual Gaussian profiles enveloping the peaks and dips.
\begin{figure}[H]
\includegraphics[width=\linewidth]{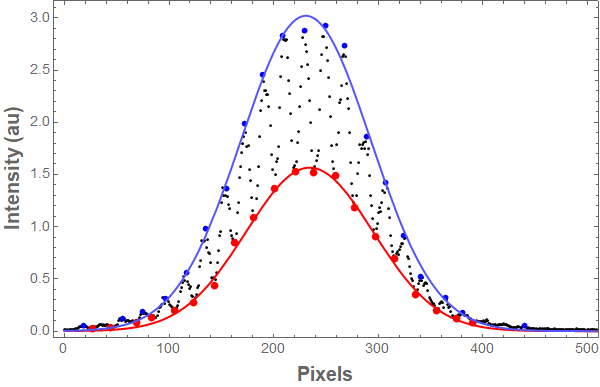}
\caption{Obtaining visibility from spatial fringes: We first find the peaks (blue dots) and dips (red dots) of the interference profile. Then we fit the peaks and dips individually with Gaussian profile $A \exp(- a (x-x_0)^2)$ to obtain the blue solid curve and red solid curve respectively. If $A_p$ and $A_d$ are the amplitudes of the Gaussian obtained from fitting the peaks and dips respectively, then the visibility is obtained by $V= \frac{A_p-A_d}{A_p+A_d}$}.
\label{vis}
\end{figure}

To measure the average value of the non-Hermitian operator $A$, we place a PBS ($R$) in one arm and a HWP($U$) in the other. To have $U = \sigma_x$, we need to have the HWP making an angle $45^ \circ$ with the Horizontal. But for completeness, the HWP is rotated from $0^\circ$ to $360 ^\circ$ in steps of $2 ^\circ$, so that we can comment on the weak value of $R$ as a function of the post-selected state $\ket{\phi(\theta)}$. $\theta$ is the angle that the  fast axis of the HWP makes with the horizontal. Although, we are terming the state $\ket{\phi}$ as post-selected state, note that there is no actual post-selection performed. Conventionally, post-selection is made after weak interaction at a later time. Here, the interaction of the beam with PBS and HWP do not have a specific time ordering and the results are independent of at which distance PBS and HWP are placed from the first beam splitter. We call $\ket{\phi}$ the post-selected state in the sense that, had we performed a conventional weak measurement of the operator $R$, we would have choosen $\ket{\phi}$ as our post-selected state.

 Also, varying the post-selected state as a function HWP angle enables us to find the expectation value of a class of operators given by
\begin{align}
\begin{pmatrix}
\cos(2 \theta) & \sin(2\theta) \\
\sin(2 \theta) & -\cos(2\theta) \\
\end{pmatrix}
\begin{pmatrix}
1 & 0 \\
0 & 0 \\
\end{pmatrix}
=
\begin{pmatrix}
\cos(2 \theta) & 0\\
\sin(2 \theta) & 0 \\
\end{pmatrix} .
\label{classOperators}
\end{align}

We record a horizontal slice of the interference pattern using a Line Camera (Thorlabs LC100M). The Line camera is placed on the exit port of the second beam splitter of the Mach Zehnder interferometer where the beam converges (see Figure \ref{fig_Coliinearity}) at a distance where the beams maximally overlap.

We then obtain visibility as described in Figure \ref{vis}. But, we could also obtain phase shift by fitting the beam profile with the following function
\begin{align}
A_0 \exp\left(-\frac{(x-\mu)^2}{2 \sigma^2}\right) 
(1 + V \cos(k x + \alpha )) .
\label{model}
\end{align}
Here, $A_0$ is the amplitude of the Gaussian envelope with the standard deviation of $\sigma$ centred at $\mu$. $k$ here represents the inverse of fringe width and x is in term of pixels.  $\alpha$ gives the phase shift of the cosine from the centre of the Gaussian envelope. If we stabilize the interferometer, we can have the zero error for $\epsilon$ fixed and then we can also determine $\varphi=\alpha+\epsilon$.  Since the interferometer was not stabilized $\alpha$ can change randomly. If the beams have slight transverse separation at the line camera, which can occur due to slight angular displacements caused when we insert or rotate optical components, then as the amplitude of two beams vary, the beam profile can have asymmetric envelope. In such cases, the model described in expression (\ref{model}) does not perfectly capture the effect of two Gaussian envelopes. Nevertheless, the visibility obtained from the model with single Gaussian model, is fairly accurate and computationally inexpensive to fit, when we have slight asymmetry.

\begin{figure}[H]
\includegraphics[width=\linewidth]{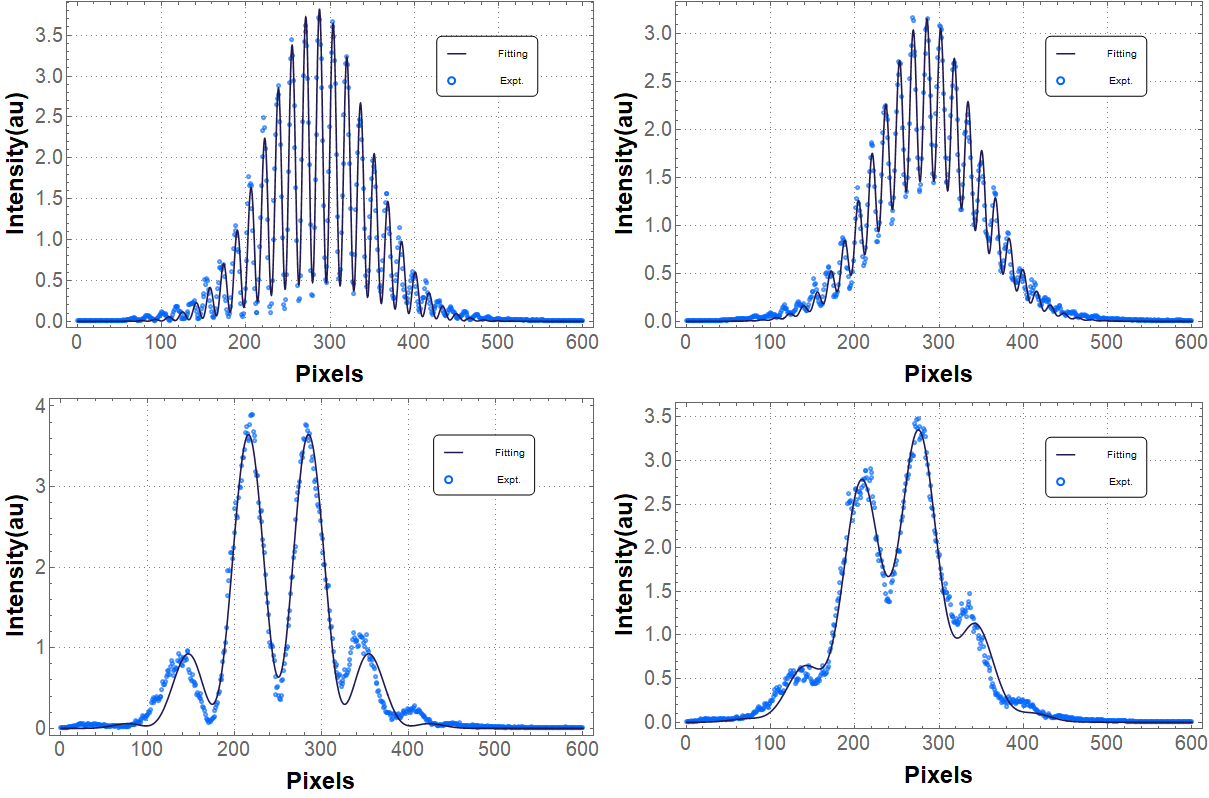}
\caption{The top row has profiles with lower fringe width compared to the bottom row. The left column has higher visibility compared to the right column.}
\label{3_fitting}
\end{figure}

Since visibility is obtained from spatial fringes, pixel size would be lowering it due to spatial averaging. If the amplitude at any position $x$ is given by $\mathcal{E}(x)$, due to pixel centred at $x$ having size $\mathcal{A}$ will record the intensity
\begin{align}
I(x) = \int_{x-\mathcal{A}}^{x+\mathcal{A}} \mathcal{E^*}(\chi) \mathcal{E}(\chi) d\chi .
\end{align}
Hence, we tweak the final beam splitter to create fringes of various width to verify that detector averaging do not affect our experiment significantly. The Figure \ref{3_fitting} below contains fringes of different visibility and fringe width.

We consider various datasets with different fringe width to verify that our results do not suffer from detector averaging. In some data sets, we also replace the Line camera with Beam profiler (WinCamD-UCD15) to vary the pixel size. In each dataset we take 100 profiles for each Half wave plate angle to eliminate random errors like fluctuation in beam profile. We fit the beam profiles and obtain the mean visibility as a function of half-wave plate angle. PBS (Thorlabs PBS 122) is also replaced with a polarizer (Thorlabs LPVIS100MP) to ensure that the results are unaffected.

\begin{figure}[H]
\includegraphics[width=\linewidth]{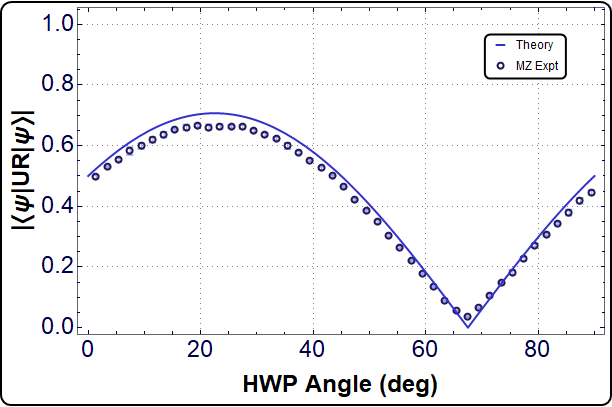}
\caption{The expectation value of a class of non-Hermitian operators described in Eq. (\ref{classOperators}) as a function of HWP angle placed in one arm of the MZ interferometer when the other arm has a polarizing beam splitter in place}
\label{4_MZ_A}
\end{figure}

After obtaining visibility $V$ for the case where we have PBS ($R$) in one arm and HWP $U(\theta)$ in the other, to obtain $|z| = \braket{\psi|UR|\psi}$, we have to multiply the factor $\frac{1+ \braket{R^2}}{2}=3/4$. The expectation value of a class of non-Hermitian operators thus obtained as a function of HWP angle is plotted in the Figure \ref{4_MZ_A}.

For HWP angle of $45 ^\circ$, we infer $\braket{A}$ from above graph to be $0.480 \pm 0.004$.

The error bars in Figure \ref{4_MZ_A} are the standard deviation of visibility over 100 profiles. These errors include the effect of any disturbance in the beam profile originating say due to mechanical vibrations in the setup. Also, imperfect beam shape or dust may also give rise to slightly off fitting results, which contributes to these error bars. The expectation value obtained in experiment do not overlap the same obtained in theory because of elliptic polarization introduced at each reflections moderates the visibility from going too high or two low.


The weak value of $R$ can be obtained by dividing $\braket{\psi|UR|\psi}$ with $\braket{\phi|\psi}$. To obtain $\braket{\phi|\psi}$, we remove the PBS so that only HWP is placed in one arm. 
\begin{figure}[H]
\includegraphics[width=\linewidth]{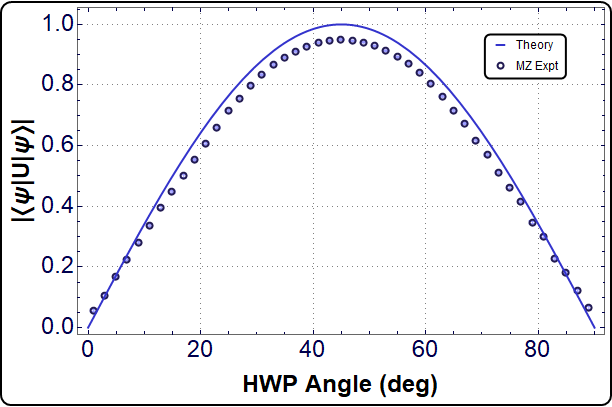}
\caption{Visibility as a function of HWP angle placed in one arm of the MZ interferometer}
\label{5_MZ_O}
\end{figure}

We obtain the weak value of $R$ from the ratio of $\braket{\psi|UR|\psi}$ obtained in Figure \ref{4_MZ_A} to visibility obtained in Figure \ref{5_MZ_O} as given in Eq. (\ref{z def}).

\begin{figure}[H]
\includegraphics[width=\linewidth]{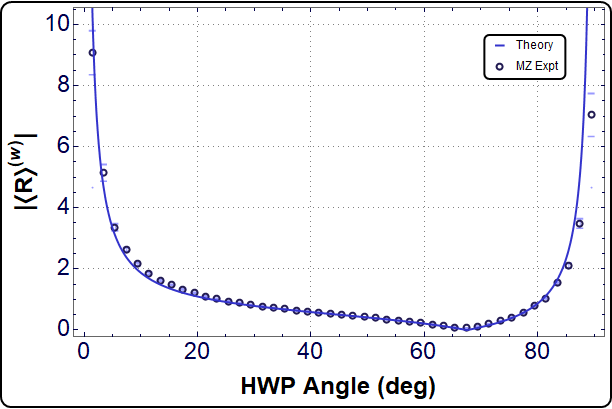}
\caption{Weak Value of $R$ inferred from MZ set up}
\label{6_MZ_R}
\end{figure}

The weak value of $R$ with pre-selected state $\ket{\psi}$ and post-selected state $\ket{\phi} = \sigma_x \ket{\psi}$ occurs at 45 $^\circ$ HWP angle. From the above graph, we infer the weak value at that angle as $0.501\pm 0.005$, very close to the theoretical value of $1/2$.

\section{Weak measurement}
Using the MZI, we obtain the weak value without having any weak interaction, i.e., there was no element that coupled the system degree of freedom (polarization) with any pointer states as per von-Neumann measurement scheme. In this section, we obtain weak value of $R$ by conventional weak measurement.

%

\begin{figure}[H]
\includegraphics[width=\linewidth]{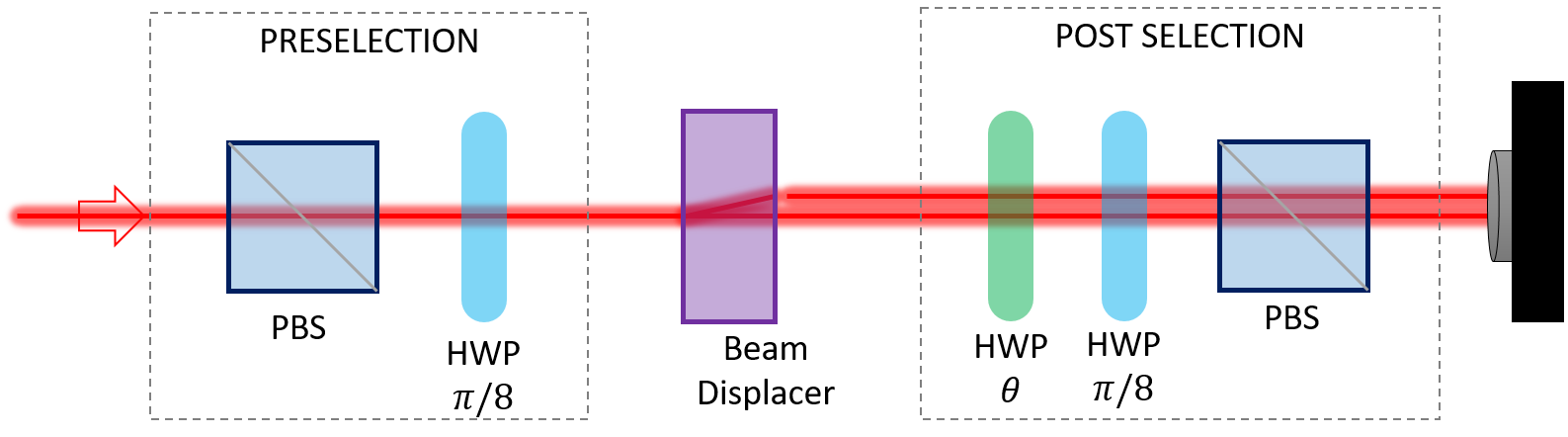}
\caption{Experimental setup to measure the weak value of $R$ with pre-selection  $\bra{+}$ and post-selected state given by $\bra{+} U(\theta)$, where $U(\theta)$ is the HWP angle that is used to vary post-selection}
\label{7_WMS}
\end{figure}

Experimentally, the preselected state is achieved by placing a half wave plate at an angle $\pi/8$ (measured from Horizontal) after a polarizing beam splitter to convert the horizontally polarized light into diagonally polarized. We vary the post-selection using a half-wave plate placed before another half-wave plate fixed at $\pi/8$  followed by a PBS. Here, we refer to the state before both the wave plates used for post-selection in Figure \ref{7_WMS}, which guarantees transmission through PBS (and hence guarantees detection with the beam profiler), as the post-selected state. 

The weak interaction occurs when the calcite beam displacer (Newlight Photonics PDC12005) lets the horizontally polarized light go undeviated and displaces the vertical component. This displacement is small compared to the standard deviation of the Gaussian beam (from He-Ne laser) which is $500 \mu m$ at the time of detection (beam divergence $< 1 m rad $). Although, its the vertically polarized beam that gets displaced, the operator description can be $R = \Pi_H $ as it is related to $\Pi_V$ by $\Pi_H = \mathds{1}-\Pi_V$  \cite{RevModPhys}. 
The weak value is experimentally obtained from the centroid shift along the direction of beam displacement normalized with the beam displacement.



\begin{figure}[H]
\includegraphics[width=\linewidth]{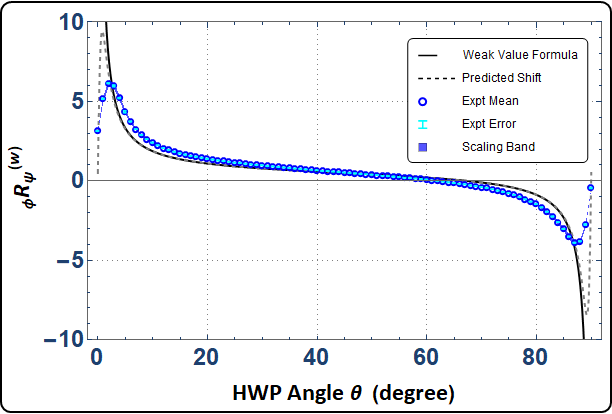}
\caption{ Absolute of weak value of $R$ in with pre-selection $\ket{+}$ and post-selection $\ket{\phi} = U(\theta) \ket{\psi}$.}
\label{wMR}
\end{figure}

The solid black line indicates the weak value of $R$ computed from Equation \ref{Eqn_weakValue} in the preselected state $\ket{\psi}$ and post-selected state $\ket{\phi(\theta)}$. The dashed line represents the weak value if we consider the more realistic $a/\sigma$ ratio. In the limit $a/\sigma \rightarrow 0$, this curve approaches the solid black line. In the experiment, at each half-wave plate angle, 10 images were obtained and the standard deviation of the centroid is represented as error bars. The experimentally obtained centroid needs to be mapped into the eigen-range and the uncertainty involved in the mapping is represented as the light blue band.

By multiplying experimentally obtained $R^{(w)}$ with theoretically known $\braket{\phi|\psi}$, we obtain the expectation value of $A$.

\begin{figure}[H]
\includegraphics[width=\linewidth]{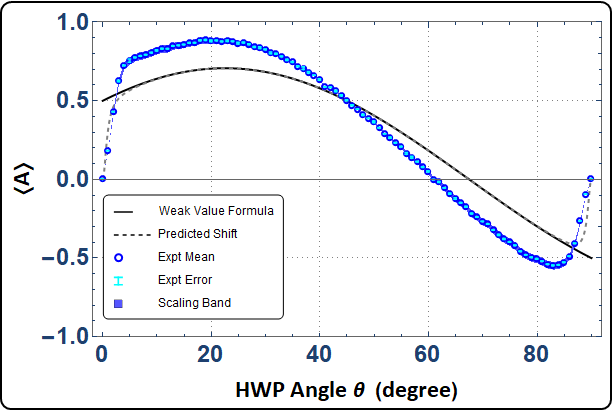}
\caption{Absolute value of Expectation value of $A$ in the state $\ket{+}$.}
\label{10_WM_A}
\end{figure}

The expectation value of $A$ obtained in this method is not robust due to slight changes in weak value. Systematic errors in centroid propagate and make the value obtained from experiment deviate from the theoretical value. We should have $\braket{A}=0.5$ at $\theta = 0$.

\section{DISCUSSION and Conclusion}

The chosen operator $A$ had a real expectation value in the given state and consequently, the weak value of $R$ turns out to be real. But in general, both the methods are well applicable to obtain complex expectation values of any non-Hermitian operator.

Theoretically, the weak value approaches infinity when pre and post-selection are orthogonal. Experimentally, this occurs only when the ratio of beam displacement to beam width tends to zero. But finite beam displacement, along with the finite extinction ratio of various polarization components make the experimentally obtained weak value finite. This when multiplied with the overlap $\braket{\phi|\psi}$, we, therefore, get zero instead of the desired expectation value of $A$. The expectation value of non-Hermitian operator for $\theta=0$ is 0.5, but from weak measurements (Figure \ref{10_WM_A}) we obtain 0. Thus, weak measurement is not a good method to infer $\braket{A}$ in the region of amplification of weak value of $R$. For all other non-Hermitian operators expressed in Eq. \ref{classOperators} parametrized by $\theta$, weak measurement provides reasonably accurate expectation value.

The MZI method circumvents the above problem as visibility directly gives a finite expectation value. However, since reflection from beam splitters/ mirrors usually introduces elliptic polarization component, additional QWP is needed to compensate for ellipticity to obtain Figure \ref{4_MZ_A}. Also, computing the weak value from visibility gives us accuracy in the amplification region (Figure \ref{6_MZ_R}).

The Mach Zehnder Interferometric method can be used to infer weak value without performing any weak measurement. The weak value obtained using weak measurement gives us information about the property of a particle at any time in between pre-selection and post-selection. Although both the experiments discussed here were performed with laser sources, the results would be the same with average statistics of single photon experiments. In the Mach Zehnder set up, however, there is no weak interaction. If the experiment is performed with single photon source, the interaction with $R$ would not necessarily occur before the interaction with $U$. In fact, the temporal ordering of interactions is irrelevant to the visibility as long as the delay is within the coherence time. Thus the MZI simulates the weak value obtained from weak measurement.


In conclusion, we have experimentally demonstrated
that expectation value of a non-Hermitian operator in any quantum state can indeed be measured. This goes beyond the traditional thinking that only Hermitian operators can be measured in experiment as they yield real eigenvalues. In fact, there are non-Hermitian operators which may also have real eigenvalues under some symmetry condition. Nevertheless, given a general non-Hermitian operator, it was not known how to measure the average of this operator in a quantum state. Remarkably, there enters the notion of weak value which allows us to measure the  average of any non-Hermitian operator by measuring the weak value of the positive semi-definite part of the non-Hermitian operator.  Even more dramatically, we have demonstrated that weak values can be experimentally obtained without performing weak measurements and without post-selection by using novel interferometric techniques. This can have several applications in measurement of weak values and
non-Hermitian operators which can have potential technological
spin-offs in future.

%
%

%
%
%
 
\begin{acknowledgments}
US would like to thank the John Templeton Foundation for funding through grant No. 57758 which enabled some of the characterizations required for this experiment. We would like to thank Sudhi Oberoi and Hafsa Syed for technical assistance.
\end{acknowledgments}

\bibliography{arXivJuly2018}


\end{document}